\documentclass[12pt]{article}
\usepackage[totalheight = 23cm, totalwidth = 17cm]{geometry}

\usepackage{amsmath, amssymb, graphics, epsfig, graphicx}
\usepackage{dcolumn}
\usepackage{bm}
\usepackage{hyperref}
\usepackage{epstopdf}

\usepackage{color}
\usepackage[section]{placeins}

\newcommand{\nc}{\newcommand}
\nc{\ba}{\begin{eqnarray}}
\nc{\ea}{\end{eqnarray}}
\newcommand\be{\begin{equation}}
\newcommand\ee{\end{equation}}

\nc{\bfk}{{\mathbf{k}}}
\nc{\bfq}{{\mathbf{q}}}
\nc{\bfp}{{\mathbf{p}}}
\nc{\bfx}{{\mathbf{x}}}
\nc{\calR}{{\cal R}}
\nc{\calP}{{\cal P}}

\nc{\eg}{{\epsilon_g}}

\nc{\cN}{{\cal N}}

\usepackage{amsmath}

\begin{document}

\begin{titlepage}

\begin{center}

\vskip 1.0cm

\large{\bf {Clustering Fossils in Solid Inflation } }

\vskip 1cm
\large{Mohammad Akhshik}
\vskip 0.5cm

\small{\it  Department of Physics, Sharif University of Technology, Tehran, Iran } \\
{m.akhshik@ipm.ir}

\vskip 1.2cm

\end{center}

\begin{abstract}

In solid inflation the single field non-Gaussianity consistency condition is violated. As a result, the long tenor perturbation induces observable clustering fossils in the form of quadrupole anisotropy in large scale structure power spectrum. In this work we revisit the bispectrum analysis for the scalar-scalar-scalar and tensor-scalar-scalar bispectrum for the general parameter space of solid. We consider the parameter space of the model in which the level of non-Gaussianity generated is consistent with the Planck constraints. Specializing to this allowed range of model parameter we calculate the quadrupole anisotropy induced from the long tensor perturbations on the power spectrum of the scalar perturbations. We argue that the imprints of clustering fossil from primordial gravitational waves on large scale structures can be detected from the future galaxy surveys. 

\end{abstract}

\end{titlepage}

\setcounter{page}{0}
\newpage
\setcounter{page}{1}

\section{Introduction}

Inflationary cosmology is the leading paradigm for providing a casual mechanism to solve the horizon and the flatness problems of the standard model of cosmology. In addition, the quantum fluctuations of
inflaton field provide the original seeds  for the fluctuations in cosmic microwave background (CMB) and the formation of large scale structure.  Simple models of inflation are based on a scalar field slowly rolling on a flat potential to yield a period of nearly 60 e-folds of inflation. These models   predict nearly scale-invariant, nearly Gaussian and nearly adiabatic perturbations which are in good agreements  with cosmological observations such as the Planck mission \cite{PLANCKInflation}.

Besides the scalar perturbations which provide the dominant contribution in CMB temperature power spectrum, the tensor perturbations are also generated during inflation. Recently, the BICEP2 experiment announced that it has found the fingerprints of the primordial gravitational waves on the CMB as these perturbations change the polarization of CMB photons and produce a significant amount of B-mode polarizations  \cite{BICEP2}. But it seems that there are unresolved issues associated with the dust and foregrounds and a detection of primordial gravitational waves  from BICEP2 mission 
is far from being conclusive   \cite{Spergel, Mortonson:2014bja}. 

At this situation, it is natural to look for other imprints of the primordial gravitational waves. As it was shown in \cite{ClusteringFossils} and further elaborated in \cite{LongTensor}, the correlation between the tensor mode and two scalars (tensor-scalar-scalar bispectrum), leads to a quadrupole anisotropy in the power spectrum of the scalar perturbations which is called ``tensor fossils''. This quadrupole anisotropy may not be detectable for the single field slow-roll  inflationary models in which the  Maldacena's  non-Gaussianity consistency relation  holds \cite{Maldacena:2002vr, Creminelli:2004yq}.
But, for the models violating this consistency relation, the quadrupole anisotropy may be large enough to be detected \cite{LongTensor}. So, in principle it is possible to rule out the standard single field slow-roll inflation with the observation of a quadrupole in scalar power spectrum.

Solid inflation \cite{Solid} is a novel model which  violates  Maldacena's consistency relation.  Recently, the tensor fossils for this model have been studied in \cite{SolidNonattractor}. But, as we will elaborate below, the parameter space of the model which was proposed in \cite{Solid} as the natural limit of the theory, which is also  used in \cite{SolidNonattractor}, leads to very large  non-Gaussianity in squeezed limit which is in tension with the Planck constraints on non-Gaussianity \cite{PLANCKNonG}. In order to ease this tension one needs to consider a new corner of solid parameter space in which the magnitude of produced non-Gaussianity is under control. However, in this new parameter space, the original results of \cite{Solid} and \cite{SQZSolid} can not be borrowed directly 
as there will be other contributions to bispectrum which are not negligible and will affect the predictions for the clustering fossils. 

In this paper, first we look in the parameter space of the solid which yields acceptable level of non-Gaussianity.  Then, for this new regime of parameter space, we calculate the tensor-scalar-scalar 
bispectrum which is subsequently used to study tensor fossils in solid inflation.

Here is the plan of the paper. In the Section \ref{solid}, we review the main aspects of solid inflation. In Section \ref{NG}, we calculate the scalar-scalar-scalar bispectrum for the solid inflation and 
look for the parameter space of the model which yields non-Gaussianity consistent with the  Planck constraints.  In Section \ref{TSS} we calculate the scalar-tensor-tensor bispectrum corresponding to this new parameter space  which will be used to  compute the tensor fossils in solid inflation in Section \ref{clustering}. The conclusions are given in Section \ref{conclusion}. 

\section{Solid Inflation}
\label{solid}

Solid inflation \cite{Solid} is an interesting inflationary model from various perspectives. In this model, the ``solid" is described by the three scalars $\phi ^I$, which provide the 
``internal coordinates" for the solid.  The  ground state of the solid is identified by 
\begin{equation}
\phi ^I =x^I, \label{Bg}
\end{equation}
in which $x ^I,  I=1, 2, 3$,  are the comoving coordinates. 

Applying this model to cosmological context, the above equation should result in an FRW background. Clearly, the equation \eqref{Bg}  breaks the isotropy and homogeneity of the background. So one has to impose further symmetries in order to obtain an isotropic and homogeneous cosmological solution.  These are,
\begin{align}
\phi ^I &\rightarrow \phi ^I +a^I, \\
\label{SO3}
\phi ^I &\rightarrow O^I_J\phi ^J,
\end{align}
in which $O$ is a member of $SO(3)$ rotation and $a^I$ is a constant.  Equipped with  these  the Lagrangian should be built out from 
\begin{equation}
B^{IJ}=g^{\mu \nu}\partial _\mu \phi ^I \partial _\nu \phi ^J,
\end{equation}
Here $B^{IJ}$ is a three by three matrix with the internal indices. Note that the capital indices $I, J,...$
are raised and lowered by the Euclidean metric $\delta_{IJ}$.  In order to impose the rotation symmetry  Eq. (\ref{SO3})  these indices should be contracted. There are three independent invariants for this matrix, which conveniently are chosen to be,
\begin{equation}
X= \left[B\right], \qquad Y=\frac{\left[ B^2\right]}{X^2}, \qquad Z=\frac{\left[B^3\right]}{X^3},
\end{equation} 
in which $\left[\quad \right]$ stands for trace. The most general action becomes \cite{Solid},
\begin{equation}
S = \int d^4 x \sqrt{-g} \left(  \frac{M_P^2}{2} R + F[X, Y, Z]  \right )
\end{equation}
in which  $M_P$ is the reduced Planck mass and 
$F(X,Y,Z)$ is an arbitrary function of the $X, Y$ and $Z$. The energy momentum tensor becomes,
\begin{equation}
T_{\mu\nu}=g_{\mu\nu}F-2\partial _\mu \phi ^I \partial _\nu \phi ^J \left[ \left(F_X-\frac{2F_YY}{X}-\frac{3F_ZZ}{X}\right)\delta ^{IJ}+\frac{2F_YB^{IJ}}{X^2}+\frac{3F_ZB^{IK}B^{KJ}}{X^3}\right],
\end{equation} 
in which $F_X$ stands for derivative of $F$ with respect to $X$, etc. 

Now the usual equations governing the background expansion are 
\begin{equation}
3M_p^2H^2=-F, \qquad 2M_p^2\dot{H}=\frac{2}{3}XF_X \, ,
\end{equation}
in which a dote indicates the derivative with respect to cosmic time $t$ and 
$H$ is the Hubble expansion rate. 
For an inflationary phase with a slowly varying Hubble parameter, we impose the slow-roll condition  
$\epsilon \ll 1$ in which $\epsilon$ is the usual slow-roll parameter 
\begin{equation}
\epsilon \equiv -\frac{\dot{H}}{H^2}=\frac{\partial \ln F}{\partial \ln X}.
\end{equation}
This means that $F(X,Y,Z)$ should have a very weak $X$-dependency. This could be obtained by imposing a further symmetry \cite{Solid}. Under the transformation,
\begin{equation}
\phi ^I \rightarrow \lambda \phi ^I,
\end{equation}
$Y$ and $Z$ remain invariant but $X$ does not. So, if one assumes that this is an approximate symmetry of the solid, then the slow-roll condition is assured. 

Turning to perturbations, as the UV pathologies of the model is the ultimate interest in what follows, then considering perturbations  in Minkowski background is enough. Then, one may expand the action and consider the behavior of perturbations $\pi ^I$ defined via,
\begin{equation}
\phi ^I=x^I+\pi ^I.
\end{equation} 
The perturbations may be decomposed into the longitudinal and the transverse components,
\begin{equation}
\pi ^I = \frac{\partial _I}{\sqrt{-\nabla ^2}}\pi _L+\pi ^I_T,
\end{equation}
in which $\partial _I \pi ^I_T=0$. The second order action for these modes is:
\begin{equation}
S_2=\int d^4x\left(-\frac{1}{3}F_XX\right)\left(\dot{\mathbf{\pi}}^2-c_T^2(\partial _i \pi _j)^2-(c_L^2-c_T^2)(\nabla \cdot \mathbf{\pi})^2\right),
\end{equation}
in which the speed of propagation of longitudinal and transverse modes are given by \cite{Solid},
\begin{equation}
c_L^2=1+\frac{2}{3}\frac{F_{XX}X^2}{F_XX}+\frac{8}{9}\frac{F_Y+F_Z}{F_XX}, \qquad c_T^2=1+\frac{2}{3}\frac{F_Y+F_Z}{F_XX}.
\end{equation}
The relation between these two speeds are,
\begin{equation}
c_T^2=\frac{3}{4}\left(1+c_L^2-\frac{2}{3}\epsilon + \frac{1}{3}\eta \right),
\end{equation}
in which  $\eta\equiv \dot \epsilon/\epsilon H$ is the second  slow-roll parameter. In order to avoid superluminal pathology one should assume \cite{Solid}
\begin{equation}
0<\left(F_Y+F_Z\right)<\frac{3}{8}\epsilon |F|. \label{Superluminal}
\end{equation}
Finally, it was shown that the theory in the whole inflationary regime is weakly-coupled \cite{Solid}.

The perturbations in FRW background is also studied in \cite{Solid}. Going to spatially flat gauge we set  all scalar parts of spatial metric to zero. With these assumptions, we may conveniently choose the metric to be, 
\begin{equation}
g_{ij}=a(t)^2\exp (h _{ij}),
\end{equation}
in which,
\begin{equation}
\partial _i h_{ij}=0, \qquad h^i_i=0.
\end{equation}
Note that with this choice, no gauge freedom is left and so the matter part is totally unconstrained,
\begin{equation}
\phi ^I=x^I+\pi ^I.
\end{equation}
 In this gauge the curvature perturbations in uniform energy density surfaces, $\zeta$, 
 is given by
\begin{equation}
\zeta = \frac{1}{3}\nabla \cdot \mathbf{\pi}.
\end{equation}  
The action for $\zeta$ is very complicated and the wave function for $\zeta$ for the Bunch-Davies vacuum is given in \cite{ANISolid}, which is,
\ba
\label{zeta-eq}
\zeta (k,\tau) = C \left(- k c_L \tau \right)^{3/2} \left( 1+ B \ln (-k c_L \tau) \right) \times
\left[  -\frac{\epsilon }{3} H_\nu^{(1)}(Q) + \frac{k \tau}{3 c_L} ( 1- \epsilon)  H_{1+\nu}^{(1)}(Q)
\right] \, ,
\ea
in which $\tau$ is the conformal time related to the comic time via the scale factor $a(t)$ as $d \tau = dt/a(t)$, $s\equiv \dot c_L/H c_L$ represents the slow change in sound speed  
and  for simplicity we have defined $Q\equiv - k c_L \tau ( 1 + s)$ and 
\ba 
-3 - 2 \epsilon - \eta + 2 s    \equiv -3 - 2 B \quad , \quad
2 \nu \equiv 3  + \eta + 5 s - 2 c_L^2 \epsilon \,  \equiv 3+ 2 A \, .
\ea
The normalization constant  $C$ is given by 
\ba
C = \frac{-i \sqrt{\pi} H}{2 M_P \sqrt{2 \epsilon k^3 c_L}} \, .
\ea
The important point to note is that on super-horizon scales, $c_L k \tau \rightarrow 0$, $\zeta$ is not frozen. Indeed, it runs logarithmically 
\ba
\zeta  \propto  (- c_L k \tau)^{-A} \left(  1+ B \ln(- c_L k \tau) \right) \quad ,  \quad 
c_L k \tau \rightarrow 0 \, .
\ea
Noting that $A, B \sim \epsilon$, we see that $\zeta$ evolves like $\epsilon \ln(-c_L k \tau)$ on super-horizon scales.

With the wave-function given by Eq. (\ref{zeta-eq}), the curvature perturbations power spectrum at the end of inflation $\tau_e$ is
\ba
\langle \zeta_{\bfk_1} \zeta_{\bfk_2}^* \rangle  =  (2\pi)^3 \delta ^3(\mathbf{k_1}+\mathbf{k_2})
P_\zeta(k_1) 
\ea
with
\ba
\label{power}
P_\zeta(k) = | \zeta_k(\tau_e)|^2 \simeq 
\frac{H^2 }{4 \epsilon c_L^5 M_P^2 k^3}
\left(1+ 2 \left ( A -B\right) N  \,     \right  )\, ,
\ea
in which $N=  -\ln (-c_L k \tau_e)$ is the number of e-folds. Noting that $A, B \sim \epsilon$ the correction from the second term above is at the order of $\epsilon N $. If  $\epsilon$ is not exponentially small, say
$\epsilon \sim 0.01$, this gives corrections of  $\lesssim 1$ in power spectrum.

It was also shown that the solid  supports a long period of anisotropic inflation \cite{ANISolid}. For further information about the anisotropic solid, the reader may refer to \cite{GWSolid} and \cite{IRSolid}. 
 
As mentioned in Introduction, solid predicts non-Gaussianity which peaks in the squeezed limit. 
The bispectrum analysis  for the general model parameter space is highly complicated. In \cite{Solid}
the authors considered the limit $F_Y \sim -F_Z \sim F $ in which the bispectrum analysis simplify significantly. In this limit, the bispectrum in Fourier space is given by (for exact definition of bispectrum see next Section) 
\begin{equation}
B_{\zeta\zeta\zeta}=\frac{3}{32}\frac{F_Y}{F}\frac{1}{\epsilon ^3c_L^2}\frac{Q(\mathbf{k_1},\mathbf{k_2},\mathbf{k_3})U(k_1,k_2,k_3)}{k_1^3k_2^3k_3^3}, \label{Bispectrumz}
\end{equation}
in which $U$ and $Q$ are functions of $\mathbf{k_i}$ which their exact form can be found in \cite{Solid}. With this bispectrum, the  amplitude of non-Gaussianity in the squeezed limit
is $f_{NL}\propto \frac{F_Y}{F}\frac{1}{\epsilon c_L^2}(1-3\cos ^2 \theta)$ in which $\theta$
is the angle between the long mode and the short modes.  However, if one takes $F_Y\sim F$ as considered in \cite{Solid} and taking $c_L^2 \simeq 1/3$ and $\epsilon \sim$ few percents, then 
the above formula yields $f_{NL} > 100$ which seems too large to be consistent with the Planck constraints on non-Gaussianity \cite{PLANCKNonG}. With these discussions in mind, one may naively extrapolate the bispectrum in the Eq. (\ref{Bispectrumz}) to the limit $F_Y \ll F$ to get $f_{NL} \sim \mathrm{few}$ in order  to be consistent with the Planck data. But, in this limit we can not trust the result in Eq. (\ref{Bispectrumz}). The reason is that in order to obtain Eq. (\ref{Bispectrumz}) many sub-leading terms have been discarded which now become important if go to the limit $F_Y, F_Z  \ll F$. Therefore,  a consistent  bispectrum analysis in the limit which $F_Y, F_Z \ll F$ is in order. 
This is the main point of this paper 
which will be applied when we calculate the tensor-scalar-scalar  bispectrum $B_{h\zeta \zeta}$ for the clustering fossils. 

In next Section we calculate the scalar-scalar-scalar bispectrum, $B_{\zeta\zeta \zeta}$, for the general parameter space  of solid and check under what conditions an acceptable amount of non-Gaussianity is generated in order to be consistent with the Planck data. As discussed above, this includes the limit $F_Y, F_Z \ll F$.  This is a consistency check of our analysis. This analysis is new and is interesting by its own right but  the reader who is not interested in scalar-scalar-scalar bispectrum  may  directly jump to Section \ref{TSS} in which we perform the analysis for the tensor-scalar-scalar  bispectrum, $B_{h\zeta \zeta}$, in this limit 
to be used for the clustering fossils.

\section{non-Gaussianity in Solid Inflation}
\label{NG}

In this section we compute the the scalar-scalar-scalar bispectrum in full parameter space 
and show that there is a parametric regime in which the non-gaussianity in solid inflation is consistent with the Planck data. 

 The leading order bispectrum analysis were performed in \cite{Solid}. We extend their analysis in different directions. We calculate the leading order bispectrum, scaling like $1/\epsilon c_L^2$, taking $F_Y/F$ and $F_Z/F$ as independent parameters  which agrees with the result obtained in \cite{Solid} in the limit $F_Y= -F_Z$. In addition, we calculate the sub-leading ${\cal O}(1)$
corrections in $f_{NL}$ parameter which show interesting structures and can be important 
observationally as we discuss below. 

To simplify the analysis in \cite{Solid} they considered the limit in which 
$|F_Y/F| \sim1, |F_Z/F| \sim 1$ subject to the upper bound  imposed in \eqref{Superluminal} 
so $F_Y = -F_Z + {\cal O}(\epsilon)$.  As a result, they only kept  the leading terms 
$|F_Y/F|= |F_Z/F| \sim  {\cal O}(1) $  in the Lagrangian  and discarded the sub-leading terms which are  at the order of  slow-roll parameter $\epsilon$.   The leading value of $f_{NL}$ in this approximation is
$f_{NL} \sim F_Y/\epsilon F \sim  1/ \epsilon$ while the corrections at the order of unity in $f_{NL}$
are discarded. This procedure has an important shortcoming. The reason is that there is no fundamental reason, such as  symmetry considerations, to ensure that  $|F_Y/F|= |F_Z/F| \sim 1$. Indeed, it is quite reasonable that both $F_Y/F$ and $F_Z/F$ are at the order of $\epsilon$ such that the upper bound is automatically satisfied. 
This is certainly the case when the dominant source of energy in $F[X, Y, Z]$
is a cosmological constant so $F$ has a very weak dependence on the fields. In addition, one can imagine the situation that either $F_Y$ or $F_Z$ vanishes (or is at the order  ${\cal O}(\epsilon^2) $).
For the sake of our discussions, suppose $F_Z=0$  and $F_Y \neq 0$. Considering the upper bound, one concludes that $F_Y/\epsilon F \sim1$. As a result, the leading term
in $f_{NL}$ from the analysis of \cite{Solid} yields $f_{NL} \sim 1$. As we shall see this is the same order as the sub-leading terms which are missing in the analysis of \cite{Solid}. Along this logic, one can also imagine the situation in which both $F_Y$ and $F_Z$ vanish (or are at the order  ${\cal O}(\epsilon^2) $). This is consistent with the upper bound  in \eqref{Superluminal}. In this limit the analysis of  \cite{Solid} yields $f_{NL} \sim \epsilon$ while the missing terms in the analysis of  \cite{Solid} are at the order of $f_{NL} \sim 1$. 
Finally, in a more technical side,  the full wave function Eq. (\ref{zeta-eq}) were not present in \cite{Solid}. Instead,  they considered the wave-function in the regions 
$ |c_L k \tau| \gtrsim \epsilon$ and $ |c_L k \tau|  \lesssim \epsilon$ separately 
and calculated the in-in integrals accordingly. As we shall see, this procedure also induces error of order unity in $f_{NL}$. 

To summarize, if one is interested in at order unity contributions in $f_{NL}$, as required from the observational constraints, then one should consider the general situation in which $F_Y$ and $F_Z$
are independent subject to upper bound constraint \eqref{Superluminal}. In addition, one also has to use the full wave function Eq. (\ref{zeta-eq}) with the slow-roll corrections implemented. 

Here, first we re-derive the analysis of \cite{Solid} for the leading order term in the bispectrum  treating  $\frac{F_Y}{F}$ and $\frac{F_Z}{F}$ as independent parameters and discard the slow-roll corrections in the wave function Eq. (\ref{zeta-eq}). The analysis with 
the sub-leading Lagrangian and the slow-roll corrected wave function are presented later on.  The  Leading order Lagrangian is
\ba
\label{lead-L}
\mathcal{L}_{\mathrm{lead}}&=&a^3H^2M_p^2 \Bigg \{ \frac{F_Y}{F}\left(-\frac{16}{27}(\partial \pi)^3+\frac{8}{9}\partial \pi \partial _i \pi ^j \partial _j \pi ^i+\frac{4}{3}\partial \pi \partial _i \pi ^j \partial _i \pi ^j -\frac{4}{3}\partial _j \pi ^i \partial _j \pi ^k \partial _k \pi ^i \right) \nonumber \\
&+&\frac{F_Z}{F}\left(-\frac{64}{81}(\partial \pi)^3+\frac{4}{3}\partial \pi \partial _i \pi ^j \partial _j \pi ^i+\frac{16}{9}\partial \pi \partial _i \pi ^j \partial _i \pi ^j-\frac{2}{9}\partial _i \pi ^j \partial _j \pi ^k \partial _k \pi ^i-2\partial _j \pi ^i \partial _j \pi ^k \partial _k \pi ^i\right) \Bigg\} \nonumber \\
\ea 
in which to simplify the notation, we have removed the subscript $L$ and denote $\pi_L$ simply by $\pi$. 

Using the standard in-in formalism \cite{Weinberg:2005vy, Chen:2010xka, Wang:2013zva}
the three-point function at the end of inflation is given by 
\begin{equation}
\langle \zeta (\tau _e)^3\rangle =-i\int _{-\infty}^{\tau _e}d\tau ^\prime \langle 0 | \left[ \zeta (\tau _e)^3,H_{int}(\tau ^\prime)\right] | 0  \rangle .
\end{equation}  
Now with the leading term Lagrangian Eq. (\ref{lead-L}), one obtains, 
\ba
\langle \zeta_{\mathbf{k_1}} \zeta_{\mathbf{k_2}} \zeta_{\mathbf{k_3}} \rangle &=& \frac{27 iM_p^2}{k_1k_2k_3} \int _{p_1p_2p_3} (2\pi)^3\delta ^3 (\mathbf{p_1}+\mathbf{p_2}+\mathbf{p_3})\left(\frac{F_Y}{F}Q_Y(\mathbf{p_1},\mathbf{p_2},\mathbf{p_3})+\frac{F_Z}{F}Q_Z(\mathbf{p_1},\mathbf{p_2},\mathbf{p_3})\right)  \nonumber \\
&& \times \int _{-\infty}^{\tau _e}d\tau ^\prime a^4(\tau ^\prime) H^2(\tau ^\prime) \big \langle 
\left[ \zeta (\tau _e,\mathbf{k_1})  \zeta (\tau _e,\mathbf{k_2})  \zeta (\tau _e,\mathbf{k_3}) , \zeta (\tau ^\prime ,\mathbf{p_1})  \zeta (\tau ^\prime ,\mathbf{p_2})   \zeta (\tau ^\prime ,\mathbf{p_3})   \right] \big \rangle  , \nonumber\\ 
\ea
where we have defined, 
\begin{align}
&Q_Y(\mathbf{p_1},\mathbf{p_2},\mathbf{p_3})\equiv-\frac{16}{27}p_1p_2p_3+\frac{20}{27}(\frac{p_1}{p_2p_3}(\mathbf{p_2}\cdot \mathbf{p_3})^2+2\mathrm{perm.})-\frac{4}{3}\frac{(\mathbf{p_2}\cdot \mathbf{p_3})(\mathbf{p_3}\cdot \mathbf{p_1})(\mathbf{p_1}\cdot \mathbf{p_2})}{p_1p_2p_3} \nonumber \\
&Q_Z(\mathbf{p_1},\mathbf{p_2},\mathbf{p_3})\equiv-\frac{55}{81}p_1p_2p_3+\frac{25}{27}(\frac{p_1}{p_2p_3}(\mathbf{p_2}\cdot \mathbf{p_3})^2+2\mathrm{perm.})-2\frac{(\mathbf{p_2}\cdot \mathbf{p_3})(\mathbf{p_3}\cdot \mathbf{p_1})(\mathbf{p_1}\cdot \mathbf{p_2})}{p_1p_2p_3}. 
\end{align}

To obtain the leading order bispectrum, we only need to consider the leading contributions in terms of the slow-roll parameters in the  wave function Eq. (\ref{zeta-eq}). The leading term in wave function
scales like $x^{5/2}H^{(1)}_{5/2}$ and the integral can be performed exactly. The leading order three-point function is obtained to be,  
\begin{align}
\langle \zeta_{\mathbf{k_1}} \zeta_{\mathbf{k_2}} \zeta_{\mathbf{k_3}} \rangle _{\mathrm{lead}}&=(2\pi)^3\delta ^3 (\mathbf{k_1}+\mathbf{k_2}+\mathbf{k_3}) \times
\nonumber \\
&\frac{3}{32}\left(\frac{H}{M_p}\right)^4\frac{1}{\epsilon ^3c_L^{12}}\frac{1}{k_1^3k_2^3k_3^3}U(k_1,k_2,k_3)\left(\frac{F_Y}{F}Q_Y(\mathbf{k_1},\mathbf{k_2},\mathbf{k_3})+\frac{F_Z}{F}Q_Z(\mathbf{k_1},\mathbf{k_2},\mathbf{k_3})\right),
\end{align}
where,
\begin{align}
U(k_1,k_2,k_3)&\equiv \frac{2}{k_1k_2k_3  K_t^3}\Big[ 3\left (k_1^6+k_2^6+k_3^6\right)+20k_1^2k_2^2k_3^2+18 \left(k_1^4k_2k_3+k_2^4k_3k_1+k_3^4k_1k_2\right)
\nonumber \\
&+ 12 \left( k_1^3k_2^3+k_2^3k_3^3+k_3^3k_1^3\right) + 9\left( k_1^5k_2+5\mathrm{perm.} \right)+12\left(k_1^4k_2^2 + 5 \mathrm{perm.}\right) \nonumber \\
&+18 \left( k_1^3k_2^2k_1+5\mathrm{perm.} \right) \Big] 
\end{align}
in which $K_t= k_1 + k_2 + k_3$. 
Note that  our results coincide with the results in \cite{Solid} in the limit $F_Y = - F_Z$
in which the combination $Q_Y$ and $Q_Z$ collapse to the shape functions $Q$ defined
in \cite{Solid} via $Q\equiv Q_Y - Q_Z$.

Now we calculate the sub-leading terms in bispectrum which were not included in the analysis of
 \cite{Solid}. There are  two types of sub-leading contributions in bispectrum. The first type comes from considering the sub-leading corrections in the Lagrangian interactions in the in-in integral contracted with the leading wave function. The second type comes from taking the slow-roll corrections in the wave function contracted with the leading Lagrangian Eq. (\ref{lead-L}) in the in-in integral.  We calculate each sub-leading terms in turn. 

Let us start with the first category, i.e. corrections in bispectrum from sub-leading Lagrangian contracted with the leading wave function. Starting with the following relations,   
\begin{equation}
F_{XX}=-\frac{a^4}{9}\epsilon F \quad,  \quad
F_{XXX}=\frac{2a^6}{27}\epsilon F \quad,  \quad (F_{XZ}+F_{XY})=O(\epsilon ^2)  \, ,
\end{equation}
and with some efforts one can show that the next to leading order Lagrangian is,
\begin{equation}
\label{L-sub}
\mathcal{L}_{\mathrm{sub}}=\epsilon a^3 M_p^2H^2\left( -{\frac{8}{27}}(\partial \pi)^3+{\frac{2}{3}}\partial \pi \partial _j \pi ^k \partial _j \pi ^k  \right). 
\end{equation}
Happily  the structure of in-in integral is the same as in the leading case. Denoting this contribution by 
$\langle \zeta ^3 (\tau _e) \rangle _{\mathrm{(1)}}$ we get:
\begin{align}
\langle \zeta ^3 (\tau _e) \rangle _{\mathrm{(1)}}=(2\pi)^3\delta ^3 (\mathbf{k_1}+\mathbf{k_2}+\mathbf{k_3}) \times 
\frac{3}{32}\left(\frac{H}{M_p}\right)^4\frac{\epsilon}{\epsilon ^3c_L^{12}}\frac{1}{k_1^3k_2^3k_3^3}U(k_1,k_2,k_3)\overline{Q}(\mathbf{k_1},\mathbf{k_2},\mathbf{k_3}),
\end{align}
in which we have introduced the new shape function $\overline{Q}(\mathbf{k_1},\mathbf{k_2},\mathbf{k_3}) $ as,
\begin{equation}
\overline{Q}(\mathbf{k_1},\mathbf{k_2},\mathbf{k_3})\equiv-{\frac{8}{27}}k_1k_2k_3+{\frac{2}{9}}\left(\frac{k_1}{k_2k_3}(\mathbf{k_2}\cdot \mathbf{k_3})^2+2\mathrm{perm.}\right).
\end{equation}

Now we calculate the corrections in bispectrum from the second category in which the leading Hamiltonian is contracted with the sub-leading corrections in the wave function. The sub-leading corrections in wave function are logarithmic terms in Eq. (\ref{zeta-eq})
coming from the term $B \ln(-c_L k \tau)$ and  the corrections in the Hankel function for the the small argument limit in which $x^{-2A} \simeq 1- 2 A \ln(x)$.  With this discussion in mind,  there are six factors of $\zeta$ in the in-in integral  in the forms of $\zeta(\tau_e)^3 \zeta(\tau')^3$. There are three equal possible ways to put the sub-leading corrections in the wave function in $\zeta(\tau_e)^3 $ and three equal possible ways to put  the sub-leading corrections in $\zeta(\tau')^3$. We denote these contributions by type (2a) and (2b) respectively.  We calculate each contributions separately.

The structure of in-in integral for the case (2a) is exactly the same as in the leading order integrals, the only exception is the additional factor  $(B-A)\ln (-c_L k\tau _e)$. Therefore, we obtain:
\ba
\langle \zeta ^3 (\tau _e) \rangle _{\mathrm{(2a)}} &=&(2\pi)^3\delta ^3 (\mathbf{k_1}+\mathbf{k_2}+\mathbf{k_3}) \times \frac{9}{32}\left(\frac{H}{M_p}\right)^4\frac{({B-A})\ln(-c_L k\tau _e)}{\epsilon ^3c_L^{12}}
\nonumber \\
&&\times \frac{1}{k_1^3k_2^3k_3^3}U(k_1,k_2,k_3)\left(\frac{F_Y}{F}Q_Y(\mathbf{k_1},\mathbf{k_2},\mathbf{k_3})+\frac{F_Z}{F}Q_Z(\mathbf{k_1},\mathbf{k_2},\mathbf{k_3})\right).
\ea

Now we calculate the contributions for the case (2b). Since  $\zeta(\tau')$ has logarithmic corrections, the structure of integral in this case is somewhat different than the previous case. The key effect to note is that
the dominant contributions in the in-in integrals come entirely from the region $\tau' \rightarrow 0 $ due to the singularity of $\ln (-\tau')$. To calculate the integrals assume $k_3 \leq k_2 \leq k_1$. Defining $x=k_3c_l\tau ^ \prime$ we obtain 
\begin{align}
\langle \zeta ^3(\tau _e) \rangle _{\mathrm{(2b)}} = i\frac{M_p^2}{H^2}\frac{27k_3^3c_L^3}{k_1k_2k_3}&\int_{p_1p_2p_3}  (2\pi)^3\delta ^3 (\mathbf{p_1}+\mathbf{p_2}+\mathbf{p_3})\left(\frac{F_Y}{F}Q_Y(\mathbf{p_i})+\frac{F_Z}{F}Q_Z(\mathbf{p_i})\right)  \nonumber \\
&\times \int _{-\infty}^{x _e}\frac{1}{x^4}dx \langle \left[\zeta ^3(\tau _e,\mathbf{k}),\zeta(\frac{p_3}{k_3}x)\zeta(\frac{p_1}{k_3}x)\zeta(\frac{p_2}{k_3}x)\right] \rangle .
\end{align}
Expanding the integrand near  $x\sim 0$ which, as discussed above,  yields the dominant contributions in the integral, we obtain 
\begin{align}
\langle \zeta ^3(\tau _e) \rangle _{\mathrm{(2b)}}=&(2\pi)^3\delta ^3(\mathbf{k_1}+\mathbf{k_2}+\mathbf{k_3})\, \frac{9}{16} \left(\frac{H}{M_p}\right)^4
\frac{\epsilon \ln(-k_3c_L\tau _e)}{\epsilon ^3c_L^{12}}
\times \nonumber \\ &
\frac{1}{k_1^3k_2^3k_3^3}\left(\frac{F_Y}{F}Q_Y(\mathbf{k_1},\mathbf{k_2},\mathbf{k_3})+\frac{F_Z}{F}Q_Z(\mathbf{k_1},\mathbf{k_2},\mathbf{k_3})\right)\overline{U}(k_1,k_2,k_3),
\end{align}
where,
\begin{equation}
\overline{U}(k_1,k_2,k_3)\equiv\left(\frac{k_1^2}{k_2k_3}+\frac{k_2^2}{k_3k_1}+\frac{k_3^2}{k_1k_2}\right).
\end{equation}

Now collecting everything, the three-point function  to first order of slow roll parameter is,
\begin{align}
\langle \zeta ^3(\tau _e) \rangle =  \langle \zeta ^3(\tau _e) \rangle_{\mathrm{lead}} +  
\langle \zeta ^3(\tau _e) \rangle_{\mathrm{(1)}} + \langle \zeta ^3(\tau _e) \rangle_{\mathrm{(2a)}} 
+\langle \zeta ^3(\tau _e) \rangle_{\mathrm{(2b)}} \, .
\end{align}
Having calculated the three-point function, we can calculate the bispectrum defined via,
\begin{align}
\label{bispectrum-def}
\langle \zeta_{\bfk_1} \zeta_{\bfk_2} \zeta_{\bfk_3} \rangle \equiv (2 \pi)^2 \delta^3(\bfk_1+ \bfk_2+ \bfk_3) B_{\zeta \zeta \zeta}( \bfk_1, \bfk_2, \bfk_3) \,  .
\end{align}
As a result, the bispectrum is obtained to be,
\begin{align}
\label{bi-B}
B_{\zeta \zeta \zeta}(\mathbf{k_1},\mathbf{k_2},\mathbf{k_3}) = \frac{3H^4}{32 M_P^4} \frac{U(k_1, k_2, k_3)}{\epsilon ^3 c_L^{12}}\frac{ Q_{eff}}{k_1^3k_2^3k_3^3} 
\end{align}
in which we have defined $Q_{eff}$ as,
\begin{align}
\label{Q-eff}
Q_{eff} &=   \epsilon  \overline{Q}(\mathbf{k_1},\mathbf{k_2},\mathbf{k_3} ) \nonumber \\ 
&+  \left(\frac{F_Y}{F} Q_Y(\mathbf{k_1},\mathbf{k_2},\mathbf{k_3} )+\frac{F_Z}{F} Q_Z(\mathbf{k_1},\mathbf{k_2},\mathbf{k_3} ) \right) \left(1+ {(B-A) \sum _i N_{k_i}} +2\epsilon {\sum _i  N_{k_i}} \frac{\overline{U}}{U}\right)  \, .
\end{align}
Here $N_{k_i} = -\ln (c_L k_i \tau_e)$ represents the number of e-folds when the mode
$k_i$ has left the horizon. Note that $Q_{eff}$ is defined such that in the limit when we neglect 
the slow-roll corrections containing $A, B$ and $\epsilon$ in Eq. (\ref{Q-eff}) and $F_Y = -F_Z$
we recover the formula for $Q_{eff}= Q\equiv Q_Y - Q_Z$ defined in \cite{Solid}. 
Note that $A- B ={\cal O}(\epsilon) $ so the terms in the last bracket in Eq. (\ref{Q-eff})
have the corrections $\epsilon N$. In the analysis yielding Eq. (\ref{Q-eff}) we have 
discarded the corrections of order $\epsilon$ while keeping corrections of order $N \epsilon$.
With $N \sim 60$, we have $\epsilon N \lesssim 1$ 
which may not be negligible.
This strategy, keeping terms of order $\epsilon N$ while neglecting terms of order $\epsilon$,
will be employed consistently in the following analysis. 

Having calculated the bispectrum, we can now calculate the amplitude of non-Gaussianity, the $f_{NL}$ parameter, defined via
\begin{equation}
f_{NL}={\frac{5}{6}}\frac{B_{\zeta \zeta \zeta}(\mathbf{k_1},\mathbf{k_2},\mathbf{k_3})}{P_\zeta(k_1) P_\zeta(k_2)+ P_\zeta(k_2)P_\zeta(k_3)+ P_\zeta(k_3) P_\zeta(k_1)} \, .
\end{equation}
Using the form of power spectrum given in Eq. (\ref{power}), we obtain, 
\begin{align}
\label{fNL-final}
f_{NL}&=\frac{5}{4\epsilon c_L^2}\frac{U(k_1,k_2,k_3)}{\left({k_1^{3}+k_2^{3}+k_3^{3}}\right)} \times \nonumber \\
&\left( 1- 2 (A  -B) \frac{k_1^3 (N_{k_2}+ N_{k_3}) + k_2^3 (N_{k_1}+ N_{k_3})+ k_3^3 (N_{k_1}+ N_{k_2})}{k_1^{3}+k_2^{3}+k_3^{3} }
  \right) Q_{eff} \, .
\end{align}
Note that the term containing the factor $(A-B)$ in the first line comes from the slow-roll corrections in the wave function as given in Eq. (\ref{zeta-eq}).  Eq. (\ref{fNL-final}) is our final result for the value of $f_{NL}$. In the approximation in which $ |F_Y| \sim |F_Z| \sim -F$ such that 
$F_Y + F_Z = - {\cal O} (\epsilon)$ and one neglects the $N \epsilon$ corrections in  Eq. (\ref{fNL-final}),  our result agrees with the result of \cite{Solid}. In our analysis, we have allowed 
for the possibility that $F_Y$ and $F_Z$ are independent parameters and also 
calculated the next sub-leading terms in $f_{NL}$ containing the corrections $\epsilon N$.

Let us look at the form of $f_{NL}$ in  the  squeezed limit $k_3 \ll k_1\sim k_2$.
In this limit, we have
\ba
\overline Q= \frac{2 k_1^2 k_3}{27} (6 \cos^2 \theta -1) \, , 
Q_Y =  \frac{4 k_1^2 k_3}{27} ( \cos^2 \theta +1)  \,  ,
Q_Z =  \frac{4 k_1^2 k_3}{81} (5- 3 \cos^2 \theta )   \,   \quad (k_3 \ll k_1\sim k_2 )
\nonumber\\
\ea
and
\ba
U= \frac{15 k_1}{k_3} \quad , \quad \overline U =  \frac{2 k_1}{k_3} \, ,
\ea
in which the angle $\theta$ is defined via $\mathbf{k_1}\cdot\mathbf{k_3}=k_1k_3\cos \theta$. 
Note that due to the triangle condition $\bfk_1 + \bfk_2 + \bfk_3=0$, we also have
$\bfk_2 . \bfk_3 = - k_1 k_3 \cos \theta$ and $\bfk_1 . \bfk_2  \simeq - k_1^2$.
Putting these formulas in Eq. (\ref{fNL-final}) yields 
\ba
\label{fNL-sq}
&&f_{NL}^{\mathrm{sq}} = {\frac{25}{36  c_L^2}}  (6 \cos ^2 \theta -1)  + 
\nonumber \\
&&~~~~~{\frac{25}{18  c_L^2}} \left[\frac{ F_Y}{F}(1+\cos ^2\theta) +\frac{F_Z}{3F}(5-3\cos ^2 \theta) \right]  
\left[ \frac{1}{\epsilon}+ \frac{4( 2 N_{k_1} + N_{k_3})}{15} + \frac{B-A  }{\epsilon}  (4 N_{k_1} + 3 N_{k_3}) \right]  \nonumber \\
\ea

Now let us look at the different terms appearing in $f_{NL}^{\mathrm{sq}}$. The first term in Eq. (\ref{fNL-sq}) is model independent in the sense that it does not depend on the form of $F(X, Y, Z)$ explicitly.
Its hidden (weak-dependence) on the  model comes via $c_L^2$. This term  
does not exists  in the analysis of \cite{Solid}. Tracing this contribution to our in-in analysis, we see that this term comes from $\overline Q$  which originates from the corrections in Lagrangian given in Eq. (\ref{L-sub}).  The terms in the second line of Eq. (\ref{fNL-sq}) have similar structures as the result in \cite{Solid}. The term containing
$1/\epsilon$ is the leading term as calculated in \cite{Solid}, while the other two terms
in the last bracket in \cite{Solid} are obtained from the corrections in the wave function
which were at the order $N \epsilon$.  Also note that, as discussed before,  we have kept $F_Y$ and $F_Z$ as independent parameters.  

Now let us look at $f_{NL}^{\mathrm{sq}} $ in different limit. First consider the limit
employed in \cite{Solid} in which $F_Y= -F_Z$. In this limit we obtain
\begin{align}
f_{NL}^{\mathrm{sq}}&= {\frac{25}{36  c_L^2}}  (6 \cos ^2 \theta -1)  +\nonumber \\ &\frac{25 F_Y}{27 Fc_L^2} \left(
 \frac{1}{\epsilon}+ \frac{4( 2 N_{k_1} + N_{k_3})}{15} + \frac{B-A  }{\epsilon}  (4 N_{k_1} + 3 N_{k_3}) \right) (3 \cos^2 \theta -1 ) \, \quad  \quad   (  F_Y= -F_Z) \, .
\end{align}
If we further assume that $F_Y/F \sim1$ as employed in \cite{Solid}, the leading term
in $f_{NL}^{\mathrm{sq}}$ is $f_{NL}^{\mathrm{sq}} \simeq \frac{25 F_Y}{27\epsilon Fc_L^2} ( 3 \cos^2 \theta -1 )$, in exact agreement with the results of \cite{Solid} and \cite{SQZSolid}. 

Now consider the $F(X)$ theory in which $F_Y=F_Z=0$. For this model, up to slow-roll corrections, $c_L^2 =1/3$ and we obtain 
\ba
f_{NL}^{\mathrm{sq}}={\frac{25}{12 }}  (6 \cos ^2 \theta -1) \quad \quad (F= F(X) ) \, .
\ea
This is a very interesting result. This indicates that for all $F(X)$ theories of solid inflation, $f_{NL}^{\mathrm{sq}}$ has a universal form. Observationally this is interesting too, since its amplitude is consistent with the upper bound from the Planck observations \cite{PLANCKNonG}. 
In addition, its shape is anisotropic which can be distinguished observationally from other local-type 
non-Gaussianities.

To summarize, in this Section we have extended the analysis of \cite{Solid} to general parameter space 
taking $F_Y/F$ and $F_Z/F$ as independent parameters. As we discussed above  we are now able to  go to the limit $F_Y/F, F_Z/F \lesssim \epsilon F$ such that Eq. (\ref{fNL-sq}) yields 
$f_{NL}^{\mathrm{sq}} \sim \mathrm{few}$  as required by Planck data.  In next Section we calculate 
$B_{h \zeta \zeta}$ for this limit of parameter space.

\section{ The Tensor-Scalar-Scalar Bispectrum}
\label{TSS}

In this section we calculate the  scalar-scalar-tensor  bispectrum. Note that, this bispectrum has been 
calculated for the original limit $F_Y \sim -F_Z \sim F$ in \cite{SQZSolid} in the squeezed limit. However, as discussed in the previous section, this limit yields large non-Gaussianity which is not consistent with the Planck observations. We would like to calculate the tensor-scalar-scalar bispectrum 
in the limit $F_Y/F \lesssim \epsilon$ such that one obtains $f_{NL}^{\mathrm{sq}} \sim \mathrm{few}$ as required by observations. 

It must be stressed that $\epsilon N <1$, since the power spectrum of curvature perturbation, \eqref{power}, receives the corrections of this order. If these corrections becomes large then the perturbation theory breaks down and one can not expand \eqref{zeta-eq}. Physically, this means that 
if inflation lasts too long then $\zeta$ receives large corrections due to evolution outside the horizon which is not favored both theoretically and observationally. This point is also stressed in \cite{SQZSolid}. Now, as it is clear from \eqref{bi-B} and \eqref{Q-eff}, one may neglect the corrections of the order $\epsilon N$ in the limit $\frac{F_Y}{F}\sim \epsilon$.  In other words, one may neglect the evolution of wave functions outside the horizon. 

The main point with this assumption is that, as it is clear from the computation of the last Section, one now only has to take care  terms coming from the sub-leading Lagrangian, i.e the first category of corrections in the last Section.  For example, if we assume that $\frac{F_Y}{F}\sim \epsilon$, then $\frac{F_Y}{F}$ will be of the order of the $F_XX$. That is what we will calculate below. Before going to the details of the analysis an important comment is in order. As emphasized in \cite{GWSolid} the dominant  contributions in bispectrum come from the matter sector and the contributions from the metric perturbations are sub-leading. This is similar to the conclusion made in \cite{Emami:2013bk, Chen:2014eua} in the context of anisotropic inflation in which it is shown that 
one can safely neglect the contributions of metric sector in anisotropic power spectrum and bispectrum. 

To calculate the bispectrum, we need the cubic Lagrangian for  the tensor-scalar-scalar interaction.  To calculate the  cubic Lagrangian first we need to calculate the perturbations in  $F(X,Y,Z)$ to third order.
Going to flat gauge, we have 
\begin{align}
F(X,Y,Z)&=F_X\delta X + F_Y \delta Y + F_Z \delta Z + \nonumber \\
& +\frac{1}{2}\left(F_{XX}\delta X^2+F_{YY}\delta Y^2+F_{ZZ}\delta Z^2+2F_{XY}\delta X \delta Y+2F_{XZ}\delta X \delta Z+2F_{YZ}\delta Y \delta Z\right). \label{expansionofF}
\end{align}
Fortunately, a great simplification will occur. We have (note that we do not distinguish between the capital index $I$ and the lower index $i$ any more)
\begin{align}
B^{ij}&=\frac{1}{a^2}\delta _{ij}-\frac{1}{a^2}h_{ij}+\frac{1}{a^2}\left(\partial _i \pi ^j +\partial _j \pi ^i\right)-\dot{\pi}^i\dot{\pi}^j \nonumber \\
&+\frac{1}{a^2}\partial _k \pi ^i \partial _k \pi ^j-\frac{1}{a^2}\left(h_{ik}\partial _k \pi ^j+h_{jk}\partial _k \pi ^i\right) -\frac{1}{a^2}h_{kl}\partial _k \pi ^i \partial _k \pi ^j.
\end{align}
We may neglect the term containing the time derivatives because it is of the $O(\epsilon ^2)$. Note that to linear order in $\pi ^i$  and $h_{ij}$, but keeping the terms of the order $O(\pi h)$, we have,
\begin{equation}
X=\frac{3}{a^2}+\frac{2}{a^2}\partial _i \pi ^i -\frac{2}{a^2}h _{ij}\partial _j \pi ^i, 
\end{equation}
and,
\begin{equation}
Y=\frac{1}{3}, \qquad Z=\frac{1}{9},
\end{equation}
which means that to this order, $Y$ and $Z$ are equal to their background values. On the other hand $\delta X$ does not contain any term of $O(h)$ since the $h_{ij}$ is traceless. With this in mind, all of the terms in the second line of \eqref{expansionofF}, except the $F_{XX}$, will not contribute to the $h\zeta \zeta$ Lagrangian. Now,
\begin{equation}
\delta X \supset -\frac{1}{a^2}h_{kj}\partial _k \pi ^i \partial _j \pi ^i=-\frac{1}{3}\overline{X}h_{kj}\partial _k \pi ^i \partial _j \pi ^i,
\end{equation}
in which $\overline{X}$ means the background value of $X$, and,
\begin{align}
\delta Y &\supset \frac{8}{9}h_{ij}\partial _i \pi ^j \partial _k \pi ^k-\frac{4}{9}h_{ij}\partial _i \pi ^k \partial _j \pi ^k-\frac{2}{9}h_{ij}\partial _k \pi ^i \partial _k \pi ^j-\frac{4}{9}h_{ij}\partial _i \pi ^k \partial _k \pi ^j, \\
\delta Z &\supset \frac{32}{27}h_{ij}\partial _i \pi ^j \partial _k \pi ^k-\frac{5}{9}h_{ij}\partial _i \pi ^k \partial _j \pi ^k-\frac{1}{3}h_{ij}\partial _k \pi ^i \partial _k \pi ^j-\frac{2}{3}h_{ij}\partial _i \pi ^k \partial _k \pi ^j. 
\end{align}
Plugging all of the pieces together, the $h\pi \pi $ Lagrangian becomes,
\begin{align}
\mathcal{L}=a^3\Bigg[&-\frac{1}{3}F_X\overline{X}h_{kj}\partial _k \pi ^i \partial _j \pi ^i-\frac{4}{9}F_{XX}\overline{X}^2h_{ij}\partial _i \pi ^j \partial _k \pi ^k + \nonumber \\
&+F_Y\left( \frac{8}{9}h_{ij}\partial _i \pi ^j \partial _k \pi ^k-\frac{4}{9}h_{ij}\partial _i \pi ^k \partial _j \pi ^k-\frac{2}{9}h_{ij}\partial _k \pi ^i \partial _k \pi ^j-\frac{4}{9}h_{ij}\partial _i \pi ^k \partial _k \pi ^j\right)+ \nonumber \\
&+F_Z\left(\frac{32}{27}h_{ij}\partial _i \pi ^j \partial _k \pi ^k-\frac{5}{9}h_{ij}\partial _i \pi ^k \partial _j \pi ^k-\frac{1}{3}h_{ij}\partial _k \pi ^i \partial _k \pi ^j-\frac{2}{3}h_{ij}\partial _i \pi ^k \partial _k \pi ^j\right)\Bigg].
\end{align} 
Note that neglecting $F_X$ and $F_{XX}$, and putting $F_Z=-F_Y$, we recover the result of \cite{SQZSolid}.

With the result of previous section in hand, we may calculate the scalar-scalar-tensor bispectrum with in-in formalism. For this purpose, we need the wave function of the tensor modes in Fourier space.
To leading order the wave function of $h$ is \cite{Solid},
\begin{equation}
\label{h-eq}
h^s_{ij}(k,\tau )=\sqrt{\frac{\pi}{2}}\frac{H}{M_pk^{3/2}}(-k\tau)^{3/2}H^{(1)}_{3/2}(-k\tau)\epsilon ^s_{ij}(k) \, ,
\end{equation}
in which $\epsilon ^s_{ij}(k)$ is the polarization tensor for the two polarizations $s =\pm$. These two polarizations are transverse to the direction of the propagation of the gravitational waves,
\begin{equation}
k^i\epsilon ^s_{ij}(\mathbf{k})=0.
\end{equation} 
Also, they satisfy the orthogonality condition,
\begin{equation}
\epsilon ^s_{ij}(\mathbf{k})\left(\epsilon ^{s^\prime {}ij}(\mathbf{k})\right)^*=2\delta ^{ss^\prime}.
\end{equation}
Now, the leading power spectrum of gravitational waves is
\begin{equation}
P_h(k)=\frac{H^2}{M_p^2}\frac{1}{k^3}. 
\label{PowerGW}
\end{equation} 

Using the standard in-in formalism, we have 
\begin{equation}
\left\langle h^s(\mathbf{k_1},\tau _e) \zeta (\mathbf{k_2},\tau _e) \zeta (\mathbf{k_3},\tau _e)\right\rangle = i \int _{-\infty} ^{\tau _e} d\tau ^\prime \left\langle 0 | h^s(\mathbf{k_1},\tau _e) \zeta (\mathbf{k_2},\tau _e) \zeta (\mathbf{k_3},\tau _e)\mathcal{L}(\tau ^\prime)|0 \right\rangle +c.c. 
\end{equation}

Before any calculation, we must Fourier transform the Lagrangian. With the relations,
\begin{equation}
\frac{F_XX}{F}=\epsilon , \qquad \frac{F_{XX}X}{F_X}=-1+O(\epsilon), \qquad 3M_p^2H^2=-F,
\end{equation}
and bearing in mind that we neglect $O(\epsilon ^2)$ corrections, and with eliminating $\pi$ in favor of $\zeta$, we get,
\begin{align}
&\big\langle h^s(\mathbf{k_1},\tau _e) \zeta (\mathbf{k_2},\tau _e) \zeta (\mathbf{k_3},\tau _e)\big\rangle \nonumber \\ &= i9 M_p^2\sum _{s^\prime}\int _{-\infty}^{\tau _e}d\tau ^\prime a^4H^2(\tau ^\prime)\int \frac{d^3\mathbf{p_1}d^3\mathbf{p_2}d^3\mathbf{p_3}}{(2\pi)^6}\delta ^3 (\mathbf{p_1}+\mathbf{p_2}+\mathbf{p_3}) \Big \langle h^{s^\prime}_{p_1}(\tau ^\prime)\zeta _{k_2}(\tau _e)\zeta _{k_3}(\tau _e)h^s_{k_1}(\tau _e) \nonumber \\
&\times \Big[ \epsilon \epsilon _{s^\prime}^{kj}\widehat{p_{2k}}\widehat{p_{2i}}\widehat{p_{3j}}\widehat{p_{3i}}\zeta _{p_2}(\tau ^\prime)\zeta _{p_3}(\tau ^\prime)-\frac{4}{3}\epsilon \epsilon _{s^\prime}^{ij}\widehat{p_{2i}}\widehat{p_{2j}}\zeta _{p_2}(\tau ^\prime)\zeta _{p_3}(\tau ^\prime) \nonumber \\
&+\frac{F_Y}{F}\epsilon _{s^\prime}^{ij}\left(-\frac{8}{3}\widehat{p_{2i}}\widehat{p_{2j}}+\frac{4}{3}\widehat{p_{2i}}\widehat{p_{2k}}\widehat{p_{3j}}\widehat{p_{3k}}+\frac{2}{3}\widehat{p_{2k}}\widehat{p_{2i}}\widehat{p_{3k}}\widehat{p_{3j}}+\frac{4}{3}\widehat{p_{2i}}\widehat{p_{2k}}\widehat{p_{3k}}\widehat{p_{3j}}\right)\zeta _{p_2}(\tau ^\prime)\zeta _{p_3}(\tau ^\prime) \nonumber \\
&+\frac{F_Z}{F}\epsilon _{s^\prime}^{ij}\left(-\frac{32}{9}\widehat{p_{2i}}\widehat{p_{2j}}+\frac{5}{3}\widehat{p_{2i}}\widehat{p_{2k}}\widehat{p_{3j}}\widehat{p_{3k}}+\widehat{p_{2k}}\widehat{p_{2i}}\widehat{p_{3k}}\widehat{p_{3j}}+2\widehat{p_{2i}}\widehat{p_{2k}}\widehat{p_{3k}}\widehat{p_{3j}}\right)\zeta _{p_2}(\tau ^\prime)\zeta _{p_3}(\tau ^\prime)\Big] \Big \rangle +c.c.
\end{align}
The factor 9  in second line comes from the relation between $\zeta$ and $\pi$ in flat slicing 
$\zeta = \frac{1}{3}\nabla \cdot \mathbf{\pi}$. 

Now we may use the standard wick theorem and the relation $aH\tau = -1 +O(\epsilon)$ to simplify the integral. We are ultimately interested in the squeezed limit, i.e $k_1=k_L << k_2\sim k_3=k_S$. However, before going to squeezed limit we comment that for a more general shape, for example the equilateral shape, the integral is dominated by a logarithmic enhancement and hence is proportional to the number of e-folds $N$. However,  in the squeezed limit this enhancement cancels out.  Now going to squeezed limit and taking into account the two possible contractions for $\zeta$, we get:
\begin{align}
\left\langle h^s(\mathbf{k_1},\tau _e) \zeta (\mathbf{k_2},\tau _e) \zeta (\mathbf{k_3},\tau _e)\right\rangle &= \nonumber \\
&\frac{18iM_p^2}{H^2}\left(-\frac{1}{3}\epsilon + \frac{2}{3}\frac{F_Y}{F}+\frac{10}{9}\frac{F_Z}{F}\right)\epsilon ^s_{ij}(\mathbf{k}_1)\hat{k}^i_2\hat{k}^j_2(2\pi)^3\delta ^3\left(\mathbf{k}_1+\mathbf{k}_2+\mathbf{k}_3\right) \nonumber \\ 
&\times \int ^{\tau _e}_{-\infty}\frac{d\tau ^\prime}{\tau ^{\prime 4}}h^s_{k1}(\tau ^\prime ) h^{s*}_{k_1}(\tau _e )\zeta _{k2}(\tau ^\prime)\zeta ^*_{k2}(\tau _e)\zeta _{k3}(\tau ^\prime)\zeta ^*_{k3}(\tau _e) + c.c
\end{align}
Now, with the wave functions for $h$ in Eq. (\ref{h-eq}) and for  $\zeta$ in Eq. (\ref{zeta-eq}) (note that we neglect the slow-roll corrections in $\zeta$ wave function) 
we may cast the integral into the following form,
\begin{align}
\left\langle h^s(\mathbf{k_1},\tau _e) \zeta (\mathbf{k_2},\tau _e) \zeta (\mathbf{k_3},\tau _e)\right\rangle &= \nonumber \\
&-\frac{\sqrt{2}}{32}\left( \frac{H}{M_p}\right)^4\frac{\pi ^{3/2}}{\epsilon ^2 c_L^{10}}\frac{1}{k_1^3k_2^6}(2\pi)^3\delta ^3\left(\mathbf{k}_1+\mathbf{k}_2+\mathbf{k}_3\right) \nonumber \\
&\times \left(-\frac{1}{3}\epsilon + \frac{2}{3}\frac{F_Y}{F}+\frac{10}{9}\frac{F_Z}{F}\right)\epsilon ^s_{ij}(\mathbf{k}_1)\hat{k}^i_2\hat{k}^j_2\left( \frac{-10\sqrt{2}}{\pi ^{3/2}}c_L^3k_2^3\right).
\end{align}
Now re-writing the above result in terms of the gravitational waves and scalar perturbations power spectra, the bispectrum $B_{h\zeta \zeta}$, defined similarly as in Eq. \ref{bispectrum-def},  
in the squeezed limit is obtained to be 
\begin{equation}
B_{h\zeta \zeta}=\frac{5}{2}P_\zeta (k_S)P_h (k_L)\frac{1}{\epsilon c_L^2}\left(-\frac{1}{3}\epsilon +\frac{2}{3}\frac{F_Y}{F}+\frac{10}{9}\frac{F_Z}{F}\right) \epsilon ^s_{ij}\hat{k}^i_S\hat{k}^j_S \, . \label{Bispectrum}
\end{equation}

Eq. (\ref{Bispectrum}) is the main result of this Section. 
Note that now $F_Y$ and $F_Z$ are independent parameters which contribute differently into
$B_{h\zeta \zeta}$. In the limit when one neglects  $\epsilon$ and letting $F_Z=-F_Y$, we recover the result of \cite{SQZSolid} and \cite{SolidNonattractor}. But now, we are allowed  to consider the new limit  $F_Y \lesssim \epsilon  F$ too, since we have already taken care of the rest of the contributions of the $O(\epsilon)$. We stress that it is not consistent to use directly the bispectrum of \cite{SQZSolid} in the limit  $F_Y \lesssim \epsilon  F$. 

\section{Clustering Fossils  in  Solid Inflation}
\label{clustering}

The effect of any field other than inflaton on the late time observables is an interesting question. This effect arises from  the coupling of this field to inflaton. This coupling may change the scalar perturbations and especially, their power spectrum. By altering the power spectrum of primordial scalar perturbations, this field can affect the late time observable too, as the scalar perturbations in inflationary era is the seed of structure formation, etc.

It was shown in  \cite{ClusteringFossils}  that the change in the primordial power spectrum induced from the long tensor mode is due to the tensor-scalar-scalar bispectrum  \cite{ClusteringFossils},
\begin{equation}
\langle\Phi (\mathbf{k_1})\Phi (\mathbf{k_2})\rangle |_{h(\mathbf{k_L})}=f(\mathbf{k_1},\mathbf{k_2}) \, h^*_p(\mathbf{k_L})\epsilon ^p _{ij}(\mathbf{k_L})k_1^ik_2^j\, 
\delta ^3 ({\mathbf{k_1}+ \mathbf{k_2}+\mathbf{k_L}}), \label{PowerFossils} 
\end{equation} 
in which  $\Phi$ stands for scalar perturbations in the $g_{00}$ component of the metric (the Newtonian potential) while the function $f(\mathbf{k_1},\mathbf{k_2})$ is given by the 
bispectrum of $h\Phi\Phi$ via
\begin{equation}
B_{h(\mathbf{k_L})\Phi(\mathbf{k_1})\Phi(\mathbf{k_2)}}=P(k_L)f(\mathbf{k_1},\mathbf{k_2})\epsilon ^p _{ij}(\mathbf{k_L})k_1^ik_2^j \, .
\end{equation}
It is clear from equation \eqref{PowerFossils} that the effect of a long tensor perturbation on a local observable is a quadrupole. But, due to the scale invariance of tensor perturbation in inflation, this quadrupole is IR divergent and becomes proportional to $N$, i.e. number of e-folds \cite{LongTensor}.

This is not the whole story. As it was shown in \cite{LongTensor}, in order to relate the quadrupole in primordial power spectrum to late time observations, there are several other steps. Authors of \cite{LongTensor} considered a galaxy survey. In order to relate this primordial power to power spectrum of galaxies, one has to track the fate of the tensor perturbation in late time universe. This mode may couple to scalar modes in late time and becomes imprinted in density perturbations of dark matter in the second order perturbation theory. On the other hand, in a galaxy survey, the location and redshift of a galaxy is inferred from a light that reaches the observer with the assumption of an unperturbed background. But, the null-geodesics of light are affected by the tensor perturbations, in an analogy with the usual lensing effect. Therefore, there is a difference between the actual position of the galaxy and the position which is presented in a galaxy survey. This is the projection effect which is  studied extensively in the literature, for example \cite{CosmicRulers} and \cite{GalaxyClustering}. As it was shown in \cite{LongTensor} for an inflationary theory in which the Maldacena's consistency relation  holds the primordial IR divergence is cancelled with the projection effect. But, a small integrated contribution remains which is called the ``tensor fossils'' \cite{LongTensor}. This fossil effect is also confirmed in the conformal Fermi normal coordinate approach of \cite{CFNC}.

Therefore, as just mentioned, the Maldacena's consistency condition plays important roles in tensor fossil effects.  In models  which violate the consistency condition, this fossil effect can be large. With this motivation in mind,  recently in  \cite{SolidNonattractor},  the tensor fossil effects in solid inflation and non-attractor models \cite{nonattractor}, as two known examples of single field models which violate the   Maldacena's consistency condition, have been studied.   The tensor-scalar-scalar bispectrum which they used is obtained with the assumption of $F_Y\sim F$, which as we already pointed out, is in  some tensions with the Planck constraints on non-Gaussianity. In order to ease the tension, the authors of \cite{SolidNonattractor} extended the results of \cite{Solid} and \cite{SQZSolid}
to the limits $F_Y \lesssim \epsilon F$. However, as we have discussed in Section \ref{NG}, in this limit there are other contributions to the bispectrum 
which can not be neglected and  are not captured in the analysis  of \cite{Solid} and \cite{SQZSolid}.  Now, with the tensor-scalar-scalar bispectrum
calculated in Eq. (\ref{Bispectrum}) valid for general values of $F_Y/F$ and $F_Z/F$ ( subject to 
the upper bound  \eqref{Superluminal}),    we are ready to  calculate the tensor fossils in solid inflation in the  regime which leads to non-Gaussianity consistent with the Planck.

As stated earlier, there will be quadrupole in power spectrum of scalars from the imprints of the 
long tensor mode. We may parametrize this quadrupole as follows:
\begin{equation}
P_\zeta (k_S)|_{h_p(k_L)}=P_\zeta (k_S)\left( 1+Q_{ij}^p(k_L)\hat{k}_i^S\hat{k}_j^S\right),
\end{equation}
in which $Q^p_{ij}(k_L)$ may be read off from the tensor-scalar-scalar bispectrum, i.e it is a manifestation of correlation between a tensor and two scalars:
\begin{equation}
Q^p_{ij}(k_L)=\frac{B_{h\zeta\zeta}(k_L,k_S,k_S)}{P_h(k_L)P_\zeta (k_S)}h_{ij}^p(k_L).
\end{equation}
Now we may expand this quadrupole in the usual basis of $Y_l^m(\mathbf{n})$, then average of the $m$. The result is \cite{SolidNonattractor},
\begin{equation}
\overline{Q^2}=\frac{8\pi}{15}\langle Q_{ij}Q^{ij}\rangle =\frac{16}{15\pi}\int _{k_L^{min}}^{k_S^{min}}k_L^2dk_L\left[\frac{B(k_L,k_S,k_S)_{h\zeta\zeta}}{P_h(k_L)P_\zeta (k_S)}\right]^2P_h(k_L),
\end{equation}
in which $k_S^{min}$ is the smallest wave number which is probed by the observations and so $k_S^{min}<H_0$. The lower limit  $k_L^{min}$ corresponds to largest wave-length tensor perturbation which is produced during inflation. 

We can apply the neat treatment of  \cite{SolidNonattractor} to our bispectrum. Let us parametrize our bispectrum as follows which is turned out to be useful when we calculate the estimator,
\begin{equation}
B_{h\zeta \zeta}(k_L,k_S,k_S)=-\frac{3}{2}\mathcal{A}P_h(k_L)P_\zeta (k_S)\epsilon _{ij}^p\hat{k}_S^i\hat{k}_S^j, \label{bi-fossil}
\end{equation}
in which,
\begin{equation}
\mathcal{A}\equiv -\frac{5}{3}\frac{1}{\epsilon c_L^2}\left( -\frac{1}{3}\epsilon +\frac{2}{3}\frac{F_Y}{F}+\frac{10}{9}\frac{F_Z}{F}\right).
\end{equation}
Now we may compute the averaged quadrupole:
\begin{equation}
\overline{Q^2}=\frac{12}{5\pi}\mathcal{A}^2\left(\frac{H}{M_p}\right)^2\ln \left(\frac{k_S^{min}}{k_L^{min}}\right),
\end{equation} 
in which the power spectrum of gravitational waves in solid is used, i.e equation \eqref{PowerGW}. Note that this quadrupole should be consistent with the essence of perturbation theory so for $k_S^{min}=H_0$, we must have,
\begin{equation}
\frac{12}{5\pi}\mathcal{A}^2\left(\frac{H}{M_p}\right)^2|\ln (k_L^{min}H_0^{-1})|<1.
\end{equation}

For the standard single field slow-roll inflation, all of the above equations hold with $\mathcal{A}=1$. The authors of  \cite{SolidNonattractor} concentrated on the parts of bispectrum violating the consistency relation, which in our notation,  corresponds to replacing $\mathcal{A}$ by $\mathcal{A}-1$ in \eqref{bi-fossil} and follow the calculation. Following \cite{LongTensor}, we can relate the primordial bispectrum to the late time observations. As shown in \cite{LongTensor}, one of the key features of the models which obey the Maldacena's consistency relation  is that if the tensor mode is infinitely long then there is no quadrupole features in the power spectrum of galaxies. In other words, the projection effect cancels out the primordial quadrupole if the tensor mode has an infinitely long wave length. This was the motivation for the authors of \cite{SolidNonattractor} in concentrating on the parts which violate the consistency relation. In our notation, if the tensor mode becomes infinitely long, then the ``observed'' quadrupole induced from the parts which violate the consistency relation is
\begin{equation}
\overline{Q^2}_{\mathrm{observed}}=\frac{12}{5\pi}(\mathcal{A}-1)^2\left(\frac{H}{M_p}\right)^2\ln \left(\frac{k_S^{min}}{k_L^{min}}\right) \, .
\end{equation} 
As we will see, $\mathcal{A}\lesssim 6$, so there may be significant quadrupole anisotropies  in galaxy surveys due to primordial gravitational waves in solid inflation. It is in direct contrast with standard single field slow-roll inflation. 

Now, in a  manner  analogous to \cite{ClusteringFossils},   
we can construct an estimator for the detection of the primordial gravitational waves from this quadrupole correction of the power spectrum . The authors of \cite{ClusteringFossils} first constructed a minimum variance estimator for the Fourier amplitude of tensor perturbations under the null hypothesis of the statistical isotropy of scalar power spectrum and calculated the noise power spectrum which is,
\begin{equation}
P^n_p(k_L)=\left[\sum _\mathbf{k_S}\frac{|f(\mathbf{k_S},\mathbf{k_L}-\mathbf{k_S})\epsilon _{ij}^p(k_L)k_S^i(k_L-k_S)^j|^2}{2VP^{tot}(k_S)P^{tot}(|\mathbf{k_L}-\mathbf{k_S}|}\right]^{-1},
\end{equation}
in which the total power spectrum is,
\begin{equation}
P^{tot}(k)=P(k)+P^n(k),
\end{equation}
which contains both the signal $P(k)$ and the noise $P^n(k)$ power spectra. In addition, the function $f(\mathbf{k_1},\mathbf{k_2})$ is defined via
\begin{equation}
B_{h\zeta\zeta}(k_L,k_1,k_2)=P_h(k_L)f(\mathbf{k_1},\mathbf{k_2})\epsilon _{ij}(k_L)k_1^ik_2^j.
\end{equation} 
Then the authors of \cite{ClusteringFossils} constructed a minimum variance estimator for the amplitude of gravitational wave with the variance
\begin{equation}
\sigma _h ^{-2}=\sum _{\mathbf{k_L},p}\frac{\left[P_h^f(k_L)\right] ^2}{2\left(P^n_p(k_L)\right]^2},
\end{equation}
in which $P_h^f(k_L)$ is defined via
\begin{equation}
P_h(k)=AP_h^f(k).
\end{equation}
For the solid model in the squeezed limit we have,
\begin{equation}
f(\mathbf{k_1},\mathbf{k_2})=-\frac{3}{2}P(k_1)k_1^{-2}\mathcal{A}.
\end{equation}
With this relation and  using $\sum _\mathbf{k}\rightarrow \frac{V}{(2\pi)^3}\int d^3\mathbf{k}$, we have,
\begin{equation}
P^n_p(k_L)=\frac{20\pi ^2}{\mathcal{A}^2k_{max}^3},
\end{equation}
where $k_{max}$ comes from the UV cut-off on momentum integral. Then, the variance of the amplitude of gravitational wave becomes,
\begin{equation}
3\sigma _h=30\pi \sqrt{3\pi}\mathcal{A}^{-2}\left(\frac{k_{max}}{k_{min}}\right)^{-3}, \label{Variance}
\end{equation}
where $k_{min}$ comes from IR cutoff on momentum integral. Note that $k_{max}$ and $k_{min}$ essentially depends on the properties of the galaxy survey under consideration. 

Now, before proceeding and talking about the detectability of signal, we have to apply the theoretical bound on $\mathcal{A}$. This bound comes from super-luminality which we have considered earlier in \eqref{Superluminal}. This bound becomes,
\begin{equation}
-\frac{F_Y}{|F|}<\frac{F_Z}{|F|}<\frac{3}{8}\epsilon -\frac{F_Y}{|F|}.
\end{equation}
Now from weak energy condition we know that $|F|=-F$. So the bound on $\mathcal{A}$ becomes,
\begin{equation}
\frac{1}{3}\epsilon-\frac{4}{9}\frac{F_Y}{|F|}<\frac{3}{5}\epsilon c_L^2\mathcal{A}<\frac{3}{4}\epsilon -\frac{4}{9}\frac{F_Y}{|F|}.
\label{bound-A}
\end{equation}

In order to proceed further  we have to have a bound on $\frac{F_Y}{|F|}$. With the Planck constraint on non-gaussianity, it is safe to assume that,
\begin{equation}
-\epsilon < \frac{F_Y}{|F|}<\epsilon .
\end{equation}
This assumption gives a non-Gaussianity of $O(\pm 1)$. There is more chance of the detection of the signal if the variance \eqref{Variance} becomes small, i.e $\mathcal{A}$ becomes large. In the best case, $\mathcal{A}\simeq 6$ and with the amplitude of tensor $A_T\simeq 2.2 \times 10^{-9}$, we obtain,
\begin{equation}
\frac{k_{max}}{k_{min}}>1550,
\end{equation}
which means that the signal is detectable at $3\sigma$ if the galaxy survey under consideration has $\frac{k_{max}}{k_{min}}>1550$. In  standard single field slow-roll inflation  $\frac{k_{max}}{k_{min}}\sim 5000$.

Note that as authors of \cite{SolidNonattractor} pointed out, the estimation of \cite{ClusteringFossils} neglects the late time effects. Including this late time effect which is thoroughly studied in \cite{LongTensor} and \cite{SolidNonattractor}, will enhance the quadrupole with a factor of $\sim 25$.

We see that  in the most optimistic case our result  for $\frac{k_{max}}{k_{min}}$  
is higher than the result of \cite{SolidNonattractor} with a factor of about 2. This means that, considering the  limit of parameter space with proper  ranges of  $\frac{F_Y}{F}$ and $\frac{F_Z}{F}$ which result in non-Gaussianity consistent with the Planck observation,  makes the signal harder to detect. However, the detection of the signal is still possible with future galaxy surveys like EUCLID or by 21 cm observations.

\section{Conclusion}
\label{conclusion}

In this paper, we studied the ``tensor fossils''  in solid inflation. Authors of \cite{SolidNonattractor} studied the same problem recently. As we argued, the original assumption about the parameter space of the solid, i.e $\frac{F_Y}{F}\sim 1$, leads to a level of non-Gaussianity in a squeezed limit which is in tensions with the results of the Planck data. In order to be consistent with the Planck observations, one needs to consider the limit  
$\frac{F_Y}{F},  \frac{F_Z}{F} \lesssim \epsilon$.

We have  computed the scalar-scalar-scalar bispectrum in full parameter space of the model 
and showed that the limit  $\frac{F_Y}{F},  \frac{F_Z}{F} \lesssim \epsilon$
leads to a level of non-Gaussianity consistent with the Planck observations. In addition,  our calculations showed clearly that in this new limit 
there are various other terms which one has to take into account in order to consistently calculate the bispectrum. 

We have calculated  the tensor-scalar-scalar bispectrum,  $B_{h \zeta \zeta}$,  beyond what is obtained in  \cite{SolidNonattractor}. Concentrating on squeezed limit, there is quadrupole anisotropy in scalar power spectrum induced by the correlation between two scalars and one tensor \cite{ClusteringFossils}. We computed this quadrupole anisotropy and showed that it may depend on the tensor modes which have left the horizon in asymptotically early times during inflation and confirmed the results of \cite{SolidNonattractor}. 

For relating the primordial quadrupole in scalar perturbation to late time observations, one has to take into account the coupling of tensor and scalar mode in non-linear evolution of perturbations and the projection effect \cite{LongTensor}. But these are the late time effects and we do not expect that anything should change in the case that primordial power spectrum is coming from the solid. So, these two effects will be completely analogous to the results obtained in \cite{LongTensor}.

Following the general path of \cite{ClusteringFossils}, we have constructed an estimator 
in order to detect the primordial gravitational waves from the solid in the late time observations such as galaxy surveys.  We have shown that  there are  corner of solid  parameter space in which: (a)- the level of non-Gaussianity in squeezed limit is consistent with the Planck data and (b)- the quadrupole signal in power spectrum of scalars is detectable by the future galaxy surveys and 21 cm observations. As the authors of \cite{LongTensor} also stressed, detecting a large quadrupole will necessarily rule out standard single field slow-roll inflation.  

\vspace{0.5 cm}
   
{\bf Acknowledgments:}
We would like to thank Hassan Firouzjahi and Razieh Emami for many insightful discussions which led to this work. Also, we would like to thank Ali Akbar Abolhasani for useful discussions about the fossil effect.


\bibliography{basename of .bib file}

\begin{thebibliography}{99}
\bibitem{PLANCKInflation} P. A. R. Ade et al., \emph{Planck 2013 results. XXII. Constraints on inflation}, [arXiv: 1303.5082  [astro-ph.CO]].

 
\bibitem{BICEP2} P. A. R. Ade et al., \emph{BICEP2 I: Detection Of B-mode Polarization at Degree Angular Scales},  Phys. Rev. Lett. 112, 241101 (2014), [arXiv:1403.3985 [astro-ph.CO]].

\bibitem{Spergel} R. Flauger, J. C. Hill, N. Spergel, \emph{Toward an Understanding of Foreground Emission in the BICEP2 Region}, [arXiv:1405.7351 [astro-ph.CO]].

\bibitem{Mortonson:2014bja} 
  M.~J.~Mortonson and U.~Seljak,
  arXiv:1405.5857 [astro-ph.CO].
  
  \bibitem{ClusteringFossils} D. Jeong, M. Kamionkowski, \emph{Clustering fossils from the early universe}, Phys. Rev. Lett. 108 (2012) 251301, [arXiv: 1203.0302 [astro-ph.CO]].

\bibitem{LongTensor} L. Dai, D. Jeong, M. Kamionkowski, \emph{Anisotropic imprint of long wave-length tensor perturbations on cosmic structure}, Phys. Rev. D88 (2013) 4, 043507, [arXiv: 1306.3985 [astro-ph.CO]].

\bibitem{Maldacena:2002vr} 
  J.~M.~Maldacena,
  JHEP {\bf 0305}, 013 (2003)
  [astro-ph/0210603].
  
\bibitem{Creminelli:2004yq} 
  P.~Creminelli and M.~Zaldarriaga,
  JCAP {\bf 0410}, 006 (2004)
  [astro-ph/0407059].



\bibitem{Solid} S. Endlich, A. Nicolis, J. Wang, \emph{Solid Inflation}, JCAP 1310 (2013) 011, [arXiv: 1210.0569 [hep-th]].

\bibitem{SolidNonattractor} E. Dimastrogiovanni, M. Fasiello, D. Jeong, M. Kamionkowski, \emph{Inflationary tensor fossils in large-scale structure}, [arXiv:1407.8204 [astro-ph.CO]].

\bibitem{PLANCKNonG} P. A. R. Ade et al., \emph{Planck 2013 Results. XXIV. Constraints on primordial non-Gaussianity}, [arXiv: 1303.5084 [astro-ph.CO]].

\bibitem{SQZSolid} S. Endlich, A. Nicolis, B. Horn, J. Wang, \emph{The sqeezed limit of solid inflation three point function}, [arXiv: 1307.8114 [hep-th]].

\bibitem{ANISolid} N. Bartolo, S. Matarrese, M. Peloso, A. Ricciardone, \emph{Anisotropy in solid inflation}, JCAP 1308 (2013) 022, [arXiv: 1306.4160 [astro-ph.CO]] .

\bibitem{GWSolid} M. Akhshik, H. Firouzjahi, R. Emami, Y. Wang, \emph{Statistical Anisotropies in gravitational waves in solid inflation}, [arXiv: 1405.4170 [astro-ph.CO]].

\bibitem{IRSolid} N. Bartolo, M. Peloso, A. Ricciardone, C. Unal, \emph{The expected anisotropy in solid inflation}, [arXiv: 1407.8053 [astro-ph.CO]].

\bibitem{Weinberg:2005vy}
  S.~Weinberg,
  Phys.\ Rev.\ D {\bf 72}, 043514 (2005)
  [hep-th/0506236].
  
\bibitem{Chen:2010xka}
  X.~Chen,
  Adv.\ Astron.\  {\bf 2010}, 638979 (2010)  [arXiv:1002.1416 [astro-ph.CO]].  
 
\bibitem{Wang:2013zva}
  Y.~Wang,
  arXiv:1303.1523 [hep-th].
  
\bibitem{Emami:2013bk} 
  R.~Emami and H.~Firouzjahi,
  ``Curvature Perturbations in Anisotropic Inflation with Symmetry Breaking,''
  JCAP {\bf 1310}, 041 (2013)
  [arXiv:1301.1219 [hep-th]].
  
\bibitem{Chen:2014eua} 
  X.~Chen, R.~Emami, H.~Firouzjahi and Y.~Wang,
  JCAP08(2014)027
  [arXiv:1404.4083 [astro-ph.CO]].





\bibitem{CosmicRulers} F. Schmidt, D. Jeong, \emph{Cosmic Rulers},  Phys. Rev. D86 (2012) 083527, [arXiv: 1204.35625[astro-ph.CO]]  .
 
\bibitem{GalaxyClustering} F. Schmidt, D. Jeong, \emph{Large-Scale Structure with Gravitational Waves I: Galaxy Clustering}, Phys. Rev. D86 (2012) 083512, [arXiv: 1205.1512 [astro-ph.CO]].

\bibitem{CFNC} F. Schmidt, E. Pajer, M. Zaldarriaga, \emph{Large scale structures and gravitional waves III: Tidal effects}, Phys. Rev. D89 (2014) 083507, [arXiv:1312.5616[astro-ph.CO]] .

\bibitem{nonattractor}
  M.~H.~Namjoo, H.~Firouzjahi and M.~Sasaki,
  Europhys.\ Lett.\  {\bf 101}, 39001 (2013)
  [arXiv:1210.3692 [astro-ph.CO]]~;
  X.~Chen, H.~Firouzjahi, M.~H.~Namjoo and M.~Sasaki,
  Europhys.\ Lett.\  {\bf 102}, 59001 (2013)
  [arXiv:1301.5699 [hep-th]]~;
  X.~Chen, H.~Firouzjahi, E.~Komatsu, M.~H.~Namjoo and M.~Sasaki,
  JCAP {\bf 1312}, 039 (2013)
  [arXiv:1308.5341 [astro-ph.CO]].



\end{thebibliography}

\end{document}